\documentclass[debug]{rmaa}

\usepackage{paralist}
\usepackage{psfrag,color}
\usepackage[latin1]{inputenc}
\newcommand{\uchii}{\hbox{UC{ }H{ }{II}{ }}}
\newcommand{\hii}{\hbox{H{ }{II}{ }}}

\newcommand\asec   {\ifmmode{^{\prime \prime}}\else{$^{\prime \prime}$}\fi}
\newcommand\pasec  {\ifmmode{{\rlap.}^{\prime \prime}}\else{${\rlap.}^{\prime \prime}$}\fi}
\newcommand\gax    {\ifmmode{_>\atop^{\sim}}\else{${_>\atop^{\sim}}$}\fi}

\newcommand\kms    {~km~s$^{-1}$}

\title{Unveiling the Hot Molecular Core in the Ultracompact \hii Region with Extended Emission G12.21--0.10}

\author{
E. de la Fuente,\altaffilmark{1}
M. A. Trinidad,\altaffilmark{2}
A. Porras,\altaffilmark{3}
C. Rodr\'iguez--Rico,\altaffilmark{2}
E. D. Araya,\altaffilmark{4}
S. Kurtz,\altaffilmark{5} 
P. Hofner,\altaffilmark{6,7}
and A. Nigoche--Netro\altaffilmark{1} } 

\altaffiltext{1}{Instituto de Astronom\'ia y Meteorolog\'ia, Departamento de F\'isica, CUCEI, Guadalajara, M\'exico.}

\altaffiltext{2}{Departamento de Astronom\'ia, Universidad de Guanajuato, Guanajuato, M\'exico.}

\altaffiltext{3}{Instituto Nacional de Astrof\'isica, \'Optica y Electr\'onica, Tonantzintla, Puebla, M\'exico.}

\altaffiltext{4}{Physics Department, Western Illinois University, Macomb, IL, USA.}

\altaffiltext{5}{Instituto de Radioastronom\'ia y Astrof\'isica, UNAM, Morelia, M\'exico.}

\altaffiltext{6}{Physics Department, New Mexico Tech, Socorro, NM, USA.}

\altaffiltext{7}{Adjunct Astronomer at the National Radio Astronomy Observatory, Socorro, NM, USA.}

\shortauthor{de la Fuente et al.}

\shorttitle{The Hot Molecular Core of G12.21--0.10}

\fulladdresses{
\item Eduardo de la Fuente and Alberto Nigoche-Netro: Instituto de Astronom\'ia y Meteorolog\'ia, Departamento de F\'isica, CUCEI, Av. Vallarta 2602, Arcos Vallarta, 44130, Guadalajara, Jalisco, M\'exico (edfuente@astro.iam.udg.mx, anigoche@gmail.com)

\item Miguel Angel Trinidad and Carlos Rodr\'iguez-Rico: Departamento de Astronom\'ia, Universidad de Guanajuato, Apdo. Postal 144, 36000 Guanajuato, M\'exico (trinidad@astro.ugto.mx, carlos@astro.ugto.mx)

\item Alicia Porras: Instituto Nacional de Astrof\'isica, \'Optica y Electr\'onica, Luis E. Erro \#~1, Tonantzintla, Puebla, M\'exico C.P. 72840 (aporras@inaoep.mx)

\item Esteban D. Araya: Physics Department, Western Illinois University, 1 University Circle, Macomb, IL 61455, USA (ed-araya@wiu.edu)

\item Stan E. Kurtz: Instituto de Radioastronom\'ia y Astrof\'isica, Universidad Nacional Aut\'onoma de M\'exico, Antigua Carretera a P\'atzcuaro \#~8701 Ex-Hda. San Jos\'e de la Huerta, Morelia, Mich., M\'exico. C. P. 58089 (s.kurtz@crya.unam.mx)

\item Peter Hofner: Physics Department, New Mexico Tech, 801 Leroy Pl., Socorro, NM 87801, USA; and National Radio Astronomy Observatory, 1003 Lopezville Road, Socorro, NM 87801, USA (phofner@nrao.edu)}

\listofauthors{E. de la Fuente, M. A. Trinidad, A. Porras, E. D. Araya, C. Rodr\'iguez-Rico, S. Kurtz, P. Hofner A., \& Nigoche-Netro}

\indexauthor{de la Fuente, E.}
\indexauthor{Trinidad, M. A.}
\indexauthor{Porras, A.}
\indexauthor{Rodr\'iguez--Rico, C.}
\indexauthor{Araya, E. D.}
\indexauthor{Kurtz, S.}
\indexauthor{Hofner, P.}
\indexauthor{Nigoche-Netro, A.}

\abstract{We present a multiwavelength study of the cometary \hii region G12.21--0.10 using the VLA and OVRO. Both radio continuum (0.3, 0.7, 2 and 3.6~cm) and spectral lines of H41$\alpha$, $^{13}$CS(2$-$1) \& (1$-$0), and NH$_{\rm 3}$(2,2) \& (4,4) observations are included. We find two 3~mm continuum peaks toward G12.21--0.10, one of them is spatially coincident with the UC~H~II region, while the other coincides spatially with a molecular clump. We also find that the 0.7, 2 and 3.6~cm continuum and H41$\alpha$ line are only detected toward the UC~H~II region, while the $^{13}$CS, and NH$_{\rm 3}$ is spatially associated with the molecular clump. Based on the morphology, kinetic temperature ($\sim$86~K), volumetric density ($\sim$1.5$\times$10$^6$~cm$^{-3}$) and linear size ($\sim$0.22~pc) of the molecular clump, we suggest this source is consistent with a hot molecular core.}

\resumen{Presentamos un estudio de la regi\'on \hii cometaria G12.21--0.10 u\-san\-do el VLA y OVRO. Se incluyen observaciones de radio continuo (0.3, 0.7, 1.3, 2 y 3.6~cm) y de las l\'ineas espectrales H41$\alpha$, $^{13}$CS(2$-$1) \& (1$-$0), NH$_{\rm 3}$(2,2) \& (4,4). Encontramos que la emisi\'on de continuo a 3~mm se divide en dos m\'aximos, de los cuales uno esta asociado con la regi\'on \hii ultra compacta (UC), mientras que el otro coincide espacialmente con un n\'ucleo molecular. La emisi\'on de continuo a 0.7, 2 and 3.6~cm y de la l\'inea H41$\alpha$ son detectadas \'unicamente hacia la regi\'on \hii UC, mientras que la emisi\'on de las l\'ineas moleculares est\'a asociada espacialmente con el n\'ucleo molecular. En base a la morfolog\'ia, temperatura cin\'etica ($\sim$86~K), densidad volum\'etrica ($\sim$1.5$\times$10$^6$~cm$^{-3}$) y tama\~no lineal ($\sim$0.22~pc) del n\'ucleo molecular, sugerimos que esta fuente es consistente con un n\'ucleo molecular caliente.}

\addkeyword{ISM: clouds}
\addkeyword{ISM: molecules}
\addkeyword{Radio lines: ISM}
\addkeyword{Radio continuum: ISM}
\addkeyword{Stars: formation}
\addkeyword{Molecular processes}

\begin{document}

\maketitle

\section{Introduction}
\label{intro}

Hot molecular cores (HMCs) are considered tobe the birthplace of high--mass stars \citep[e.g.,][]{K00, GL99}. Thus, to study their physical properties, dynamics, and evolution is crucial for advancing our understanding of high--mass star formation. HMCs are detected mainly by emission of high--excitation molecular lines (e.g., CH$_3$CN and NH$_{\rm 3}$) and by millimeter continuum emission from warm dust. A detailed overview of star formation is presented by \citet{SP05}, while a review of high--mass star formation was published by \citet{ZY07}. 

Water maser emission is a common phenomenon in the neighborhood of newly-formed high--mass stars \citep[e.g.,][]{C04, T03, HC96}. However, the relationship of the H$_2$O masers to the star formation environment is still not fully understood. It is accepted that their presence implies the existence of at least one embedded YSO \citep[e.g.,][]{GL99}. For ultra compact \hii regions (\hbox{UC{ }H{ }{II}}) with cometary morphologies, \citet{HC96} showed that the water masers almost always located in clusters, near, yet offset, from the cometary arc; and for other morphologies the masers are often projected onto the ionized gas. \citet{HC96} also found that water masers are related to hot NH$_{\rm 3}$ clumps (and YSOs) rather than to the ionized gas of the \hii regions. The latter was also observed by \citet{C94} in a sample of 4 molecular cores. 

G12.21--0.10 (G12 hereafter), located at a distance of 13.5~kpc \citep{C90}, has been classified as a cometary \uchii region with extended emission \citep{F09a, F07, KK01}. C$^{17}$O observations finding a molecular clump coincident with the UC emission were reported by \citet{Ho00}. This region also presents emission of hot and dense chemically enriched gas tracers (see Table~\ref{previous}), and has been also observed in infrared (IR) surveys \citep{F09b, F07, D03}. Hence, it was deemed a prime candidate for a HMC by \citet{D03}.

In G12, H$_2$O and CH$_{\rm 3}$OH maser emission is observed toward the molecular clump, no towards the ionized gas \citep{K04}. In particular, the water masers are offset by at $\sim$4$\asec$ (or 0.26 pc at 13.5 kpc) from the \uchii region, suggesting the location of a deeply embedded, young high--mass star \citep{HC96}. In addition, \citet{K04} found that a strong water maser feature in this cluster coincides spatially with both class~I (44 GHz) and class~II (6.67 GHz) CH$_{\rm 3}$OH masers. Thus, G12 is an interesting candidate for studying the nature and formation of water and methanol masers and for understanding their relation to molecular gas and high--mass star formation. 

In \S~\ref{obs} we describe the radio continuum and spectral line observations and data reduction. In \S~\ref{results} we present the observational results and analysis as follows: 1.- we show that the $^{13}$CS (1$-$0 and  2$-$1) transitions trace the hot and optically thin dust clump in G12; 2.- we calculate physical characteristics of this clump using NH$_{\rm 3}$ data confirm that it is a hot molecular core. In \S~\ref{disc} we discuss the $^{13}$CS and NH$_{\rm 3}$ data in the context of the HMC interpretation. A summary is presented in \S~\ref{summary}.

\section{Observations and Data Reduction}
\label{obs}

\subsection{Owen's Valley Radio Observatory (OVRO) Observations}
\label{ovro}

The $^{13}$CS(2$-$1) and H41$\alpha$ line observations were made with OVRO during March and April of 1996. The equatorial and high resolution configurations were used, providing baselines from 30 to 240~m. Cryogenically cooled SIS receivers provided typical SSB system temperatures of 220--350~K.  The digital correlator was split into two bands to observe the $^{13}$CS(2$-$1) ($\nu_0 = 92.49430$ GHz) and H41$\alpha$ ($\nu_0 = 92.03445$ GHz) lines with 30~MHz ($\sim$100\kms) and 60~MHz ($\sim$200\kms) bandwidths, and 62 and 60 Hanning--smoothed channels, respectively. This configuration gives a spectral resolution of 0.5~MHz (1.6\kms) for the $^{13}$CS(2$-$1) line and 1~MHz (3.3\kms) for the H41$\alpha$ line. 

In addition, simultaneous continuum observations at 3~mm were made using a 1~GHz bandwidth analog correlator. 
The quasars NRAO~530, 3C 454.3, B1908$-$202, and B1749+096 were observed to track amplitude and phase variations. The bandpass calibration and flux scale were established by observing 3C 273. The absolute flux uncertainty is estimated to be $\sim$20~\%, and the initial calibration was carried out using the Caltech MMA data reduction package \citep{S93}. Further data reduction and mapping were done with the AIPS software package. The continuum obtained from the line--free channels was subtracted from the $^{13}$CS and H41$\alpha$  emission lines in the $(u,v)$ plane using the program UVLIN in AIPS. Finally, Gaussian fits to the observed spectra were made with the CLASS software package.

\subsection{Karl G. Jansky Very Large Array (VLA) Observations}
\label{vla}

\subsubsection{Radio--Continuum Emission}
\label{RCE}

High angular resolution continuum observations at 0.7, 2, and 3.6~cm toward G12 were made with the VLA\footnote{The National Radio Astronomy Observatory is a facility of the National Science Foundation operated under  cooperative agreement by Associated Universities, Inc.} in its CnB configuration on 1996 January 29. Two 50~MHz bands were observed, each one including both right and left circular polarizations. We used a subarray of 12 antennas for 7~mm observations and the remaining 15 antennas were employed for 2 and 3.6~cm observations. The flux and phase calibrators were 3C286 (1.45, 3.40 and 5.23 Jy at 0.7, 2, and 3.6 cm) and B1923+210 (1.0, 1.8 and 1.0 Jy; flux densities meassured at 0.7, 2, and 3.6 cm respectively).

In addition, low angular resolution VLA observations at 3.6~cm in the D configuration toward G12 were made on 2005 April 03. As before two 50~MHz bands were observed, each one including both right and left circular polarizations. The flux and phase calibrators were 3C 286 (5.23 Jy) and J1832$-$105 (1.28~Jy), respectively. 

The calibration and imaging for both high and low angular resolution observations were made following  standard procedures in AIPS. Self--calibration in phase only, and Multi--Resolution Cleaning were performed for the low angular resolution data.

\subsubsection{NH$_{\rm 3}$ and $^{13}$CS(1$-$0) Emission}
\label{nh3cs10}

Observations of NH$_{\rm 3}$ were made with the VLA in its DnC configuration on 2004 June 06. Both transitions (2,2) and (4,4), with rest frequencies of  23722.631 and 24139.417 MHz, respectively, were observed. In order to include the main line and one pair of hyperfine satellites, the spectral line correlator was split into two overlapping bands of 64 channels each, centered at 2.30 and 0.76 MHz higher and lower than the expected line frequency for (2,2) and 1.06 and 1.24 MHz for (4,4). The Doppler tracking center velocity was set to 24\kms ~(V$_{\rm LSR}$) and the total velocity coverage was 94\kms, obtaining a frequency resolution of 97.656 kHz (1.25\kms). The flux, phase and bandpass calibrators were 3C 286 (2.6 Jy), J1820$-$254 (0.53~Jy) and J1924$-$292 (9.1~Jy), respectively. Calibration and data reduction were done following standard line procedures of the AIPS software package. All images were obtained using the task IMAGR with ROBUST~=~0, while the spectra were obtained with task ISPEC after combining the two overlapping bands into a single 71 channel cube and the cube moment maps were made using the task XMOM. Gaussian fits to the observed spectra were made using the CLASS software package. 

During the same run, $^{\rm 13}$CS(1$-$0) observations at the rest frequency of 46.24757 GHz, were also carried out. A central bandpass velocity of 24\kms   \, was also used. Flux, phase and bandpass calibrators were the same as for NH$_{\rm 3}$, as well as the calibration and data reduction. A natural weighting was used to generate the images.

\section{Observational Results and Analysis}
\label{results}

\subsection{Radio Continuum Emission}
\label{resultsRCE}

\begin{figure}[!h]
  \includegraphics[width=\columnwidth]{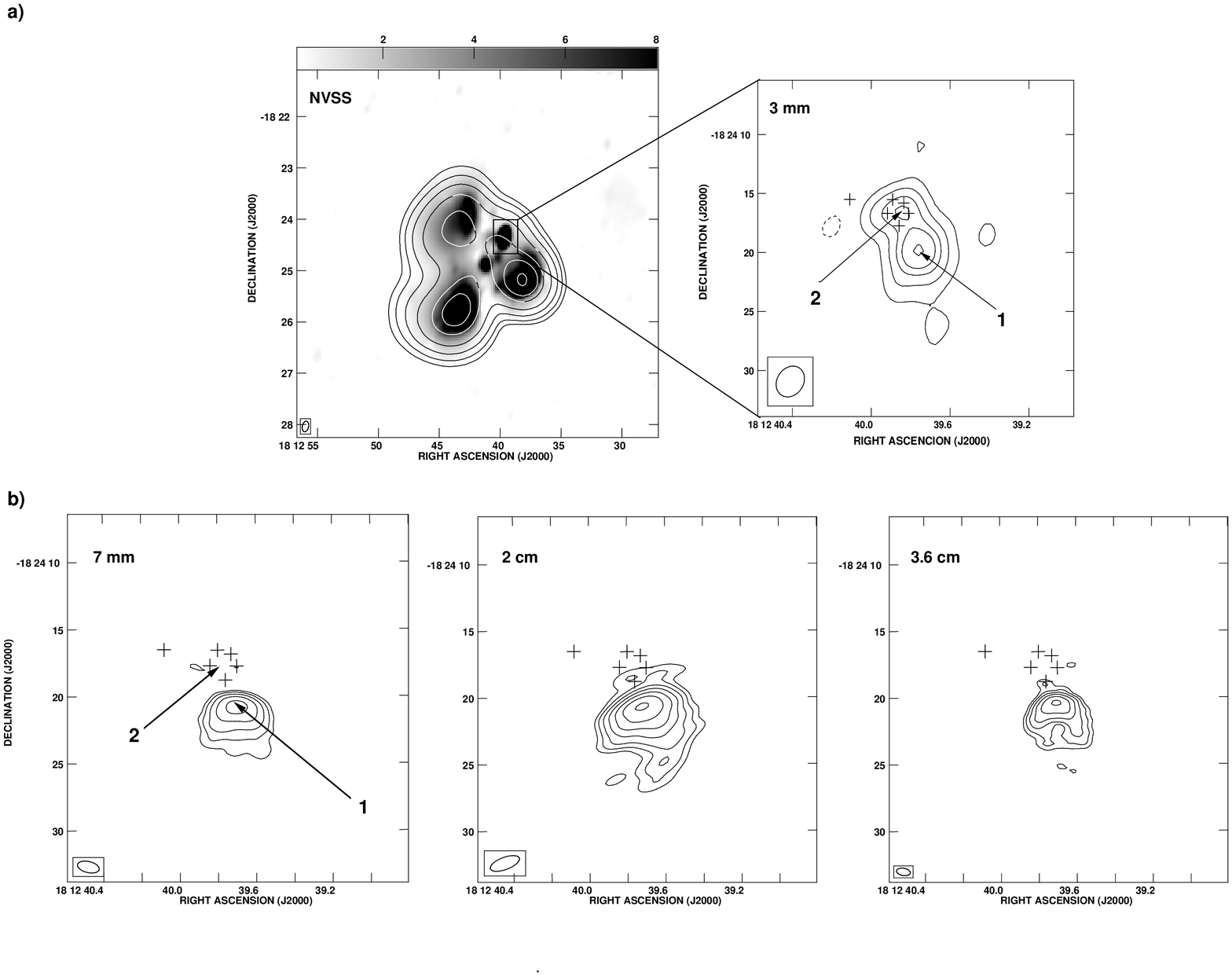}

 \caption{G12.21--0.10 (G12). {\bf a)} {\it Left}: 21 cm NVSS map (contours) superposed with 3.6~cm VLA D-configuration image in gray-scale \citep{F07}. The NVSS beam is 45$''$, while the 3.6~cm beam, shown in the lower-left corner, is $12\rlap{.}{''}5$ $\times$ $7\rlap{.}{''}3$ at P.A.~=~$-$13$^{\circ}$. Contours are --3, 3, 6, 12, 24, 48, 68, 96 $\times$  3~mJy~beam$^{\rm -1}$. The gray-scale is 0.16 to 8~mJy beam$^{\rm -1}$.  These maps represent the extended emission in G12. {\it Right}: A close up view of the UC component traced by radio continuum emission at 3~mm (see \S~2). Contours are --3, 3, 9, 18, 27, 48 times RMS noise of 1.7~mJy~beam$^{\rm -1}$. The beam size is shown in the bottom left. The positions of the the \uchii region and the molecular clump are labeled as 1 and 2 respectively. In all panels, the crosses mark the position of the water masers from \citet{HC96}. {\bf b)} Radio continuum emission at 0.7, 2, and 3.6~cm (VLA-CnB) toward G12. The contour levels correspond to --1, 1, 2, 4, 8, 16, 32, 64 $\times$ RMS noise presented in Table~\ref{vla_cont} for each wavelength. The synthesized beam for each map is shown in the lower left corner (see Table~\ref{vla_cont}). }
  \label{fig1}
\end{figure}

The 21~cm NVSS contour map \citep{Co98} superposed on the 3.6~cm low angular resolution VLA--D map in gray--scale toward G12 is shown in Figure~\ref{fig1}a~{\it Left}. The morphology of both emissions match at different scales in this region, and are in agreement with 21~cm observations from \citet[their Figure~1g]{KK01}. Six continuum peaks are detected at 3.6~cm, one of which corresponds to the UC source.

The 3~mm continuum emission toward the UC source is split into two peaks (see Figure~\ref{fig1}a~{\it Right}). They are separated $\sim$4$''$ (0.26~pc at a distance of 13.5~kpc), one of them corresponds to the \uchii region (labeled as 1), while the other (labeled as 2) coincides with a water maser group reported by \citet{HC96}. Due to the possible relation between these masers and the molecular clump reported by \citet{Ha00}, we will refer to peak--2 as the molecular clump. The peak positions of the 3~mm continuum emission from the \uchii region and the molecular clump are reported in Table~\ref{pos}. 

High angular resolution continuum observations at 0.7, 2, and 3.6~cm are shown in Figure~\ref{fig1}b. They show that the radio continuum emission is only associated with the \uchii region. Then, the observed emission at 3~mm and the lack of emission at 0.7, 2 and 3.6~cm toward the 3~mm peak--2, suggests that the 3~mm continuum emission toward the molecular clump could arise from dust rather than ionized gas. Integrated continuum flux densities of the ionized gas, rms noise, beam, and deconvolved sizes for these wavelengths are given in Table~\ref{vla_cont}. 

\subsection{Line Emission}
\label{le}

\subsubsection{H41$\alpha$}
\label{h41a}

The H41$\alpha$ ($\sim$3.3~mm) emission is only detected toward the \uchii region and no emission is detected toward the molecular clump. The H41$\alpha$ emission has a similar distribution as the radio continuum observed at 7~mm shown in Figure~\ref{fig1}b. 

The H41$\alpha$ spectrum toward G12 is shown in Figure~\ref{fig2} (upper panel). Line emission was detected in the velocity range of $\sim$10 to 45\kms, with a peak at about 30\kms. The results of the Gaussian fit of this recombination line are presented in Table~\ref{h41cspar}. 

The H41$\alpha$ velocity is similar to that of the molecular emission from  $^{13}$CO(1$-$0), C$^{34}$S(2$-$1), and C$^{34}$S(3$-$2) of 20$-$25\kms, both from \citet{KK03} and this work (see below). In addition, the H41$\alpha$ LSR central velocity is in agreement with that reported by \citet{KK01} using H76$\alpha$ for three positions in the extended emission ($\sim$27\kms ~on average, see their Figure~1g). The ionized gas in G12, present at different scales (UC and extended emission; see Figure~\ref{fig1}), coexists with the molecular gas and originates in the same star forming region.

\subsubsection{$^{\rm 13}$CS(2$-$1)}
\label{13cs21}

\begin{figure}[!h]
  \includegraphics[width=\columnwidth]{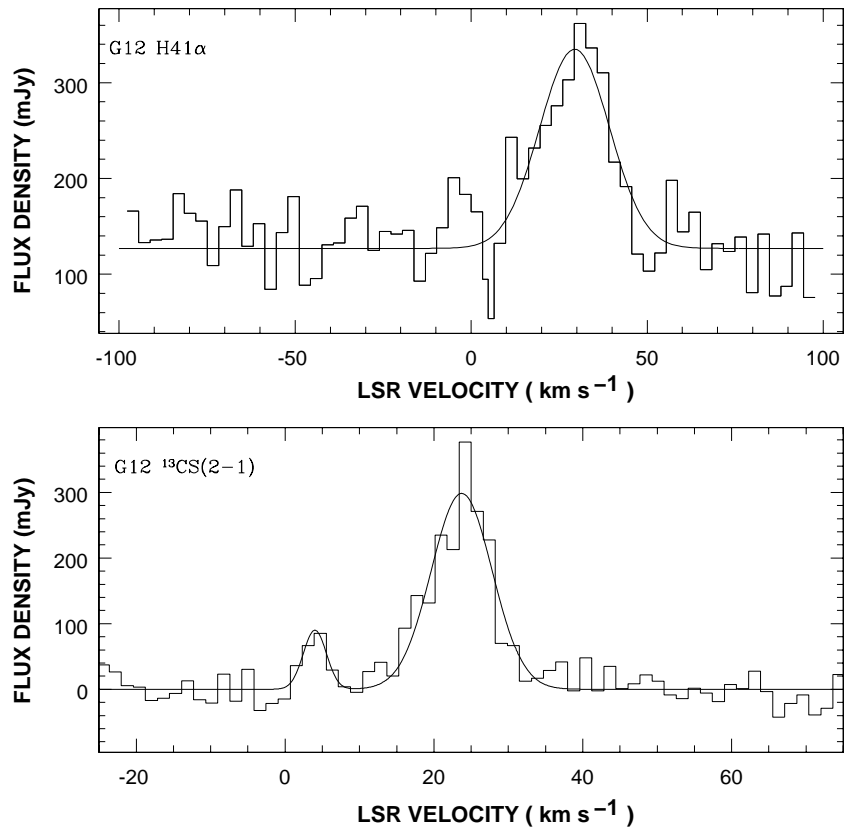}
 \caption{H41$\alpha$ and $^{\rm 13}$CS(2$-$1) spectral lines. The solid line represents the best adjusted Gaussian in each spectrum. The CS spectrum also shows an unexpected weaker line at $\sim$4.5\kms. A possible candidate for this line is C$_{\rm 3}$S.}
  \label{fig2}
\end{figure}

\begin{figure}[!h]
  \includegraphics[width=\columnwidth]{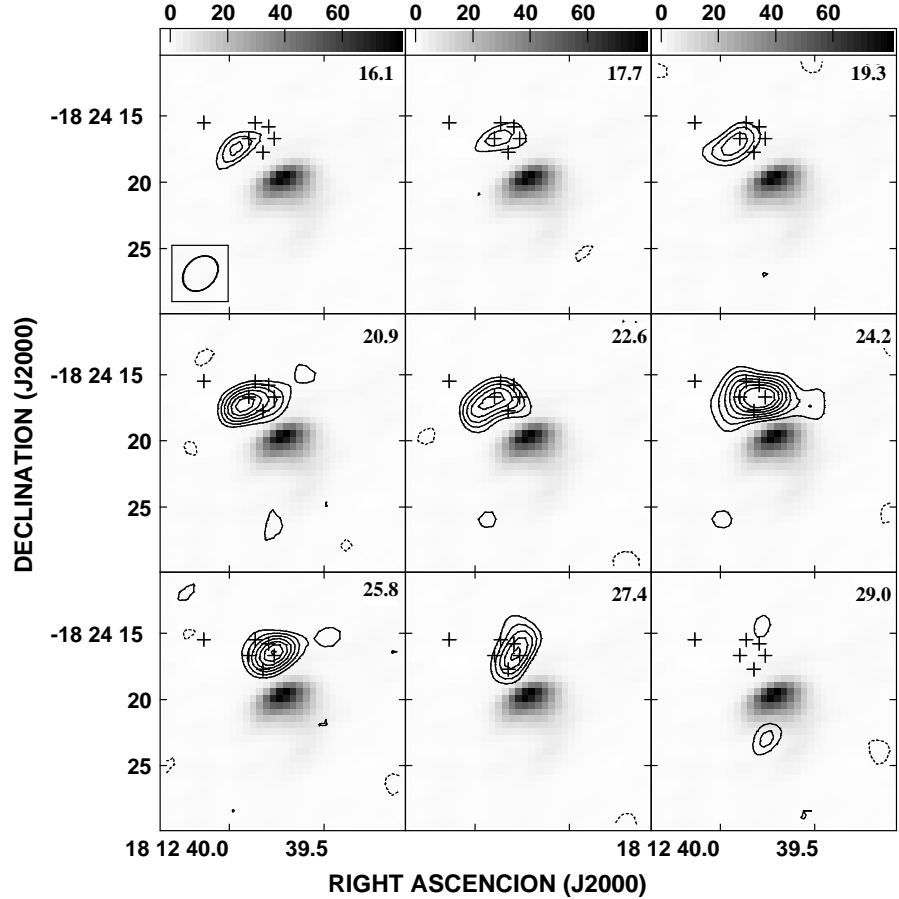}
 \caption{$^{\rm 13}$CS(2$-$1) channel maps (contours) with channel width of 1.62\kms ~are shown between 16 and 29.0\kms ~superposed on the radio continuum emission at 3.6~cm (VLA-CnB, gray-scale) of G12. The gray-scale goes from $-$1.24 to 79.83~mJy~beam$^{-1}$. Contours are --3, 3, 4, 5, 6, 7, 8, 9 $\times$ 35~mJy~beam$^{-1}$ (RMS noise). The synthesized beam of the $^{\rm 13}$CS(2$-$1) observations ($3\rlap{.}{''}18$ $\times$ $2\rlap{.}{''}51$ with a P.A. of $-$43.3$^\circ$) is shown in top left panel. The crosses mark the position of water masers from \citet{HC96}. }
  \label{fig3}
\end{figure}

Figure~\ref{fig3} shows the channel maps of the $^{13}$CS(2$-$1) line emission superposed on the 3.6~cm continuum emission.  The deconvolved size of this molecular region is $\sim 4\pasec 5 \times 2\pasec 0$ (0.36 $\times$ 0.16 pc) with the major axis oriented approximately in the E$-$W direction, while the $^{13}$CS(2$-$1) peak emission observed at V$_{\rm LSR}$~=~24\kms ~is blushifted ($\sim$5\kms) with respect to the H41$\alpha$. The $^{13}$CS(2$-$1) line spectrum is shown in Figure~\ref{fig2} (lower panel). Observed parameters (flux densities, rms noise, beam, and deconvolved sizes) for $^{\rm 13}$CS(2$-$1) toward G12 are presented in Table~\ref{h41cspar}.

Comparing the $^{13}$CS(2$-$1) line and the 3.6~cm (VLA--CnB) continuum emission, we note that the $^{13}$CS(2$-$1) emission is not spatially associated with the UC region in G12, but rather is associated is with the molecular clump and the group of water masers. This result suggests the presence of a molecular clump spatially associated with the 3~mm continuum emission (peak--2) not far from the \uchii region (see Table~\ref{pos}).

Figure~\ref{fig2} (lower panel), show that in addition to the $^{13}$CS(2$-$1) line at V$_{\rm LSR}$ = 24\kms, a weaker component at 92.49284 GHz is marginally detected at V$_{\rm LSR}$~=~4.5$\pm$0.5\kms assuming the $^{13}$CS(2$-$1) rest frequency; and ${\rm \Delta}$V~=~3.6$\pm$1.1\kms. A possible candidate to explain this weak emission is the molecule C$_{\rm 3}$S \citep[$\nu_0$ = 92.48849 GHz;][]{Mo03} at V$_{\rm LSR}$~=~14.1$\pm$0.5\kms. Although this molecular line seems to be real, given the low signal--to--noise ($\sim$2.5$\sigma$ at the peak), we cannot determine the physical parameters of the molecular gas from which this emission arises. 

\subsubsection{$^{\rm 13}$CS(1$-$0)}
\label{13cs10}

\begin{figure}[!h]
  \includegraphics[width=\columnwidth]{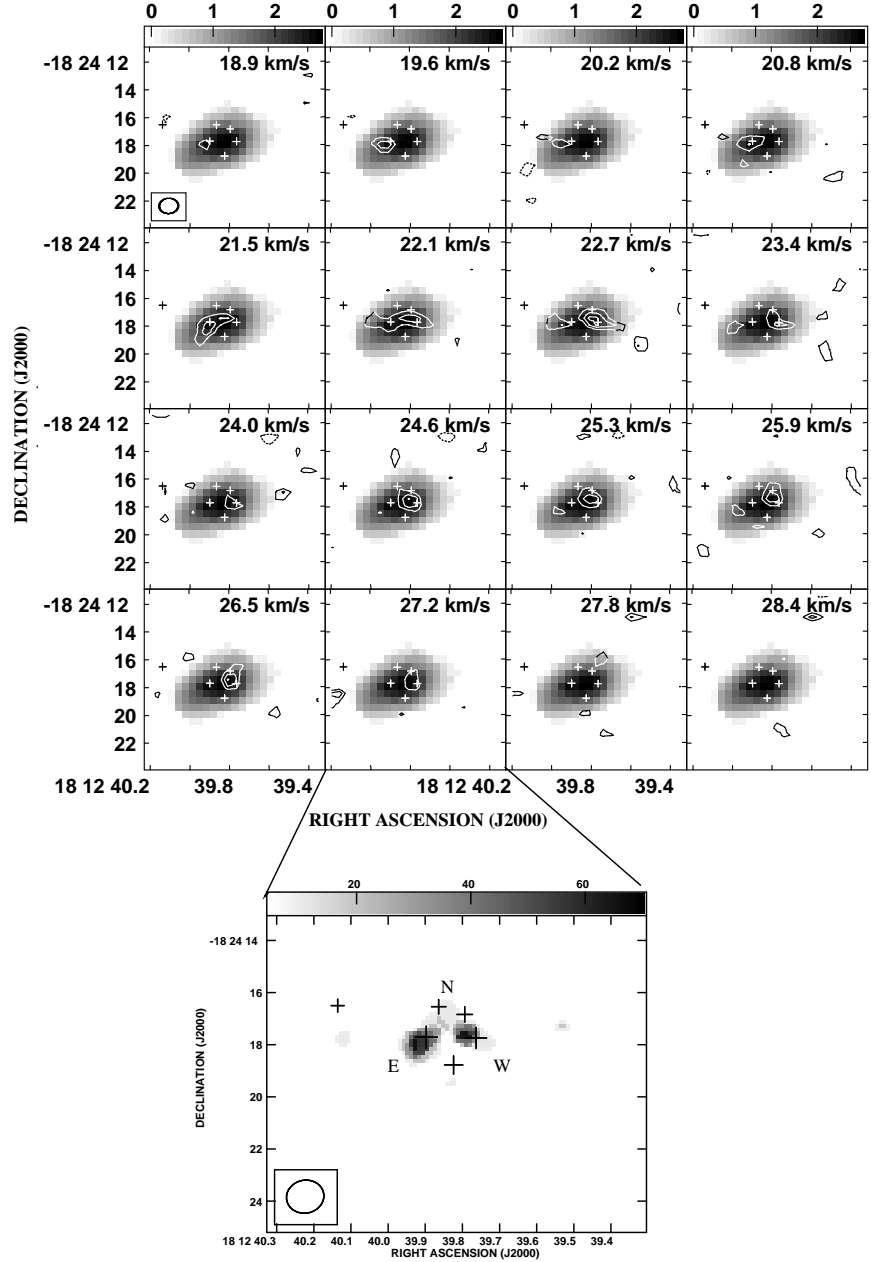}
 \caption{ $^{\rm 13}$CS(1$-$0) maps of individual velocity channels from the main line spectra of G12 (contours), superposed with the $^{\rm 13}$CS(2$-$1) integrated map (in gray scale). Velocities from 18.9 to 28.4 \kms ~are shown in each map. The $^{\rm 13}$CS(1$-$0) integrated emission map (inset in the 27.2 \kms panel) shows a complex edge-like structure morphology with three components labeled E~(East), W~(West), and N~(North). The maximum of emission for each of these components corresponds to $\sim$20, 23 and 26 \kms ~respectively. The contours are --3, 3, 4, 5 $\times$ 3.6~mJy (RMS noise). The beam size is $1\rlap{.}{''}44$ $\times$ $1\rlap{.}{''}26$ with a P.A. of --79$^\circ$, shown in the bottom left of the first map. The crosses mark the position of water masers from \citet{HC96}.}
   \label{fig4}
\end{figure}

Figure~\ref{fig4} shows the channel maps of the $^{13}$CS(1$-$0) emission towards G12 superposed to the $^{13}$CS(2$-$1) integrated map as comparison. Although $^{13}$CS(1$-$0) and $^{13}$CS(2$-$1) present different morphologies, both are spatially associated with the molecular clump (3~mm peak--2). This morphological difference could in part be due to the different spatial resolution achieved for each molecular line. However, the $^{13}$CS(1$-$0) emission presents a complex morphology as revealed by each channel map (see also the integrated emission shown in the inset of the 27.2\kms ~panel). This morphology shows a central double--component roughly in the East--West direction (labeled as E and W on the inset) with peak line emission  observed at V$_{\rm LSR}$~=~20\kms ~and~23\kms, respectively. Another weak emission peak to the North (labeled as N) is also observed at V$_{\rm LSR}$~=~26\kms.

\subsubsection{NH$_{\rm 3}$}
\label{nh3}

\begin{figure}[!h]
\includegraphics[width=\columnwidth]{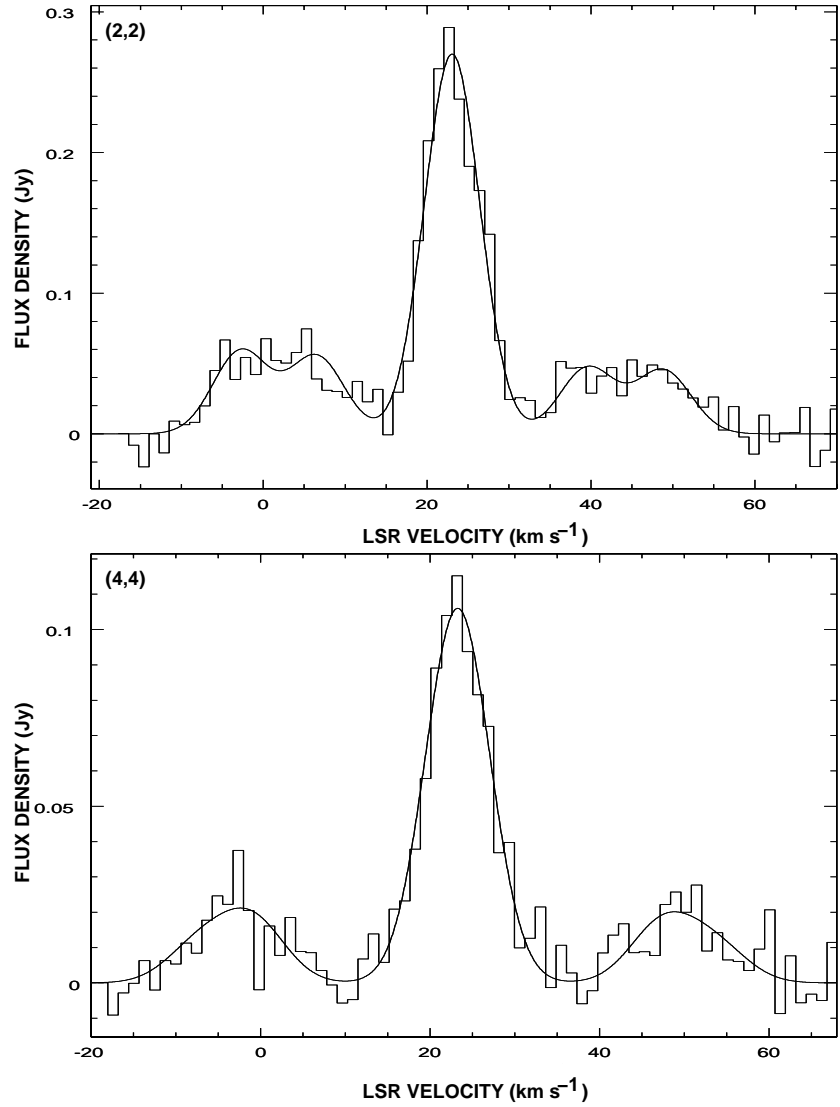}
  \caption{NH$_{\rm 3}$ integrated spectra of (2,2) and (4,4) transitions in the G12--HMC . The solid lines show Gaussian fits.} 
 \label{fig5}
\end{figure}

The NH$_{\rm 3}$ emission is only detected toward the molecular clump (3~mm peak--2). The integrated NH$_{\rm 3}$ spectra of the (2,2) and (4,4) transitions are shown in Figure~\ref{fig5}. In the (2,2) transition the satellite lines are marginally resolved, and in (4,4) the satellite lines are blended into a single component. The two electric hyperfine components were fit with two Gaussians and the results of the fits are given in Table~\ref{nh3obs}. 

The (4,4) main peak line flux density is weaker than that measured in the (2,2) line. The observed widths of 8--9\kms ~are evidence of highly supersonic turbulent motions in the molecular gas \citep{GL99}. The observed line parameters, flux densities, rms noise, beam, and deconvolved sizes for both transitions (2,2) and (4,4) are presented in Table~\ref{nh3obs}. Within the uncertainties, the (2,2) and (4,4) V$_{\rm LSR}$ and $\Delta {\rm V_{obs}}$ are in mutual agreement. The measured LSR velocities of both transitions are slightly smaller ($\sim$1\kms) than those reported by \citet{C90} and \citet{C92} for (2,2) and (4,4) using the Effelsberg 100~m telescope, while the $\Delta {\rm V_{obs}}$ for (2,2) and (4,4) are $\sim$1\kms ~and $\sim$4\kms ~greater, respectively. The line width difference indicates motion of the hotter compact (4,4) gas relative to the cooler extended (2,2) gas as discussed in \citet{C92}. 

Neither the (2,2) nor the (4,4) spectra show detectable anomalies in the satellite hyperfine components like those observed toward S106 in the NH$_{\rm 3}$ (1,1) transition \citep{S82}. Hence we can assume that the populations in the metastable levels are in LTE. To calculate physical parameters of the region based on the NH$_{\rm 3}$ spectra, we also assume that true intrinsic widths, $\Delta {\rm V_{t}}$, are equal to the observed values, $\Delta {\rm V_{obs}}$. 

Optical depths for the main lines, ${\rm \tau_{main}}$, of 3.5$\pm$0.5 and 8.0$\pm$0.5 for (2,2) and (4,4) transitions, respectively, were calculated numerically \citep[][equation 1]{HT83}. The total optical depth can be calculated as ${\rm \tau_{tot}}$~=~(1/b)${\rm \tau_{main}}$ where b represents the normalized (to total intensity) relative intensities of the main line. Parameter b values of 0.8 and 0.93 for (2,2) and (4,4) respectively are assumed \citep{H77}, obtaining $\tau_{\rm tot}^{\rm 2,2}$~=~4.4$\pm$0.5 and $\tau_{\rm tot}^{\rm 4,4}$~=~8.6$\pm$0.5. In addition, excitation temperatures (T$_{\rm exc}$) of 78$\pm$5~K and 35$\pm$5~K for (2,2) and (4,4) were calculated in a standard way by solving the transfer equation \citep{H77, HT83}. The column densities for (2,2) and (4,4) transitions in terms of ${\rm \tau_{\rm main}}$, T$_{\rm exc}$ and $\Delta {\rm V_{t}}$ were computed through the ${\rm \tau_{tot}}$ definition given by \citet{U86} using the respective Einstein coefficient for spontaneous de--excitation: A$_{\pm}$(2,2)~=~2.23$\times$10$^{-7}$~s$^{-1}$ and A$_{\pm}$(4,4)~=~2.82$\times$10$^{-7}$~s$^{-1}$. Hence N(2,2)~=~2.6$\pm$1.0$\times$10$^{16}$~cm$^{-2}$ and N(4,4)~=~2.0$\pm$1.0$\times$10$^{16}$~cm$^{-2}$. 

The rotational temperature, T$_{\rm{42}}$ = 77$\pm$10~K, was computed in terms of the (2,2) and (4,4) peak flux densities and velocity widths (see Table~\ref{nh3obs}) considering the same solid angle for both emissions. Using this rotational temperature, a rotation level diagram for para--NH$_{\rm 3}$ \citep{WU83, PK75} considering collisional transitions between K--ladders and transitions which follow the parity transition rule, and doing the calculations in statistical equilibrium derived by \citet{WU83} with the numerical computations of \citet{D88} in detailed balance, we obtain a value for the kinetic temperature of ${\rm {T_k}}$~=~86$\pm$12~K.

\section{Discussion}
\label{disc}

\subsection{The Hot Molecular Core}
\label{g12hmc}

The kinetic temperature (86~K), volumetric density (1.5$\times$10$^6$~cm$^{-3}$) and linear size ($\sim$0.22~pc, assuming a distance of 13.5~kpc) generally coincide with the operational definition of HMC; hot ($\gtrsim$~100~K), dense ($\sim$10$^5$ to 10$^8$~cm$^{\rm -3}$), and small (size~$\lesssim$~0.1~pc) molecular gas structures \citep{C92, GL99}. So, based on these characteristics, we suggest that the molecular gas clump detected in G12 is a Hot Molecular Core (G12--HMC hereafter). 

\begin{figure}[!h]
\includegraphics[width=\columnwidth]{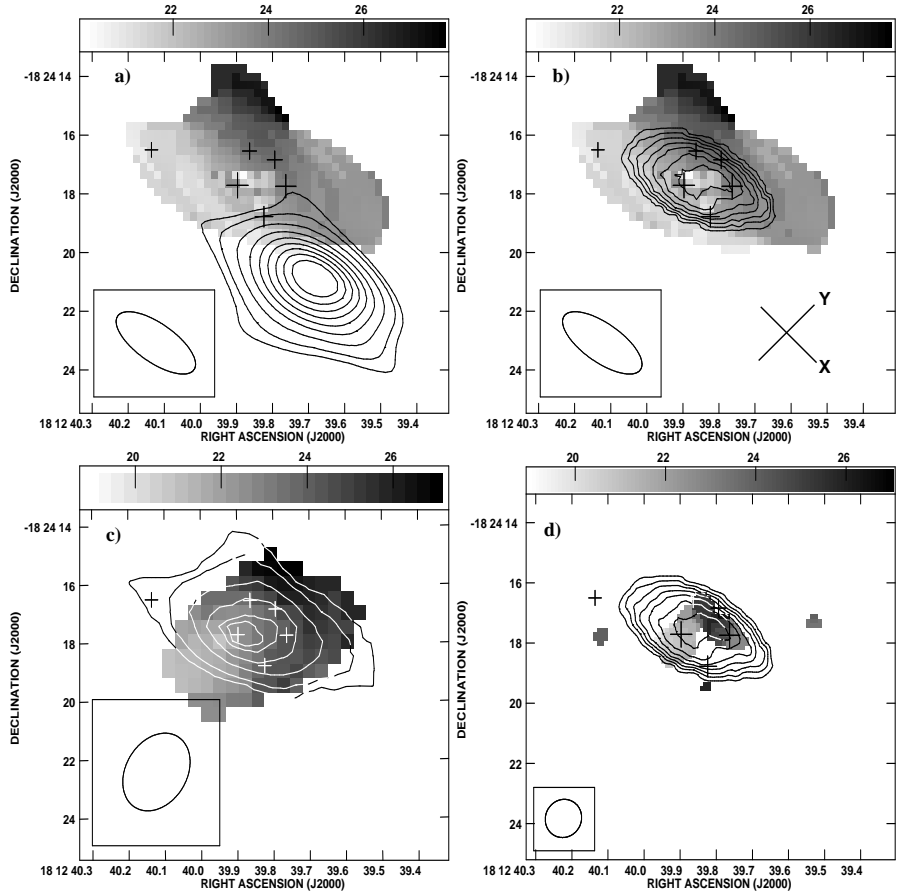}
  \caption{G12--HMC. {\bf a)} NH$_{\rm 3}$(2,2) moment--1 map (gray-scale), superposed to the radio continuum emission at 1.3~cm (contours; \uchii region) obtained from line--free channels of the NH$_{\rm 3}$(2,2) observations. Gray--scale is in the interval $\sim$ 19--27\kms. Contours are --4, 4, 6, 10, 15, 20, 25, 30, 35, 40 $\times$ 2~mJy~beam$^{-1}$ (RMS noise). {\bf b)} NH$_{\rm 3}$(2,2) moment--1 map (gray-scale) superposed to the NH$_{\rm 3}$(4,4) integrated emission (contours).  Contours are 20, 50, 100, 140, 180, 220, 260~mJy~beam$^{-1}$km~s$^{-1}$. Axes labeled as X and Y at the lower right corner show the directions used to generate the PV diagrams presented in Figure~\ref{fig7}. {\bf c)} $^{\rm 13}$CS (2$-$1) moment--1 map (gray-scale) of the emission presented in Figure~\ref{fig4}, superposed to the NH$_{\rm 3}$(2,2) integrated emission with countours in 50, 150, 250, 350, 450, 550~mJy~beam$^{-1}$km~s$^{-1}$. {\bf d)} $^{\rm 13}$CS (1$-$0) moment--1 map (gray-scale), superposed is the NH$_{\rm 3}$(4,4) integrated emission in contours as in b). Synthesized beams are shown at the bottom left of each plot and corresponds to NH$_{\rm 3}$ in a) and b), $^{\rm 13}$CS(2$-$1) in c), and $^{\rm 13}$CS (1$-$0) in d). In all figures, crosses mark the position of water masers \citep{HC96}.}
 \label{fig6}
\end{figure}

Several maps of the molecular line emission from the G12--HMC region are presented in Figure~\ref{fig6}. No radio continuum source was detected toward the G12--HMC position (Figure~\ref{fig6}a) where the group of H$_2$O masers is located. This could be explained following the suggestion of \citet{W03} that 6.67~GHz CH$_{\rm 3}$OH  masers appear before an \hii region is created. However, higher sensitivity observations are needed, since \citet{R16} have a 100\% detection rate in the radio continuum toward 25 HMCs, when observed with much higher sensitivity of $\sim$3--10~$\mu$Jy beam$^{-1}$ rms.

The NH$_{\rm 3}$(2,2) and (4,4) emission clearly coincides with the molecular clump observed at 3~mm (see Figure~\ref{fig1}a), as well as the group of masers. The distribution of molecular gas from both transitions, NH$_{\rm 3}$(2,2) and (4,4), is similar, although the NH$_{\rm 3}$(4,4) emission is slightly more compact than the NH$_{\rm 3}$(2,2) emission, suggesting that the hotter and denser gas is confined in a smaller--scale structure. Note that Figures~\ref{fig6}c and \ref{fig6}d show the more extended and more compact emission, respectively.

The first moment map of the (2,2) transition shows a velocity gradient from E to N (following the nomenclature of the inset in Figure~\ref{fig4}) from 21 to 28\kms ~over $\sim$4$\asec$ ($\sim$30~km s$^{-1}$~pc$^{-1}$). A similar behavior is observed for the (4,4) moment--1 map \citep{Fu10}, but from 21 to 27\kms ~over $\sim$3$\asec$ corresponding to 25~km~s$^{-1}$~pc$^{-1}$. For $^{13}$CS(1$-$0), from $\sim$20 to $\sim$26\kms, we obtain a velocity gradient of $\sim$3\kms\,arcsec$^{-1}$ (or 45\kms\,pc$^{-1}$), and for the $^{13}$CS(2$-$1) emission $\sim$28\kms\,pc$^{-1}$, which is similar to that of the NH$_{\rm 3}$(2,2) map (compare the grey--scale bars in Figures~\ref{fig6}a and \ref{fig6}c). The $^{\rm 13}$CS and NH$_{\rm 3}$ emission coincides in position and kinematics, which  suggests no chemical differentiation in the HMC. 

In addition to the velocity gradient from E to N for $^{13}$CS(1$-$0), a velocity gradient from W to N of $\sim$1.5\kms\, arcsec$^{-1}$ (or 23\kms\,pc$^{-1}$) is also estimated. A comparison of these results with those reported in Orion--IRc2 HMCs \citep{CW97} suggests that $^{13}$CS(1$-$0) is a good candidate line to trace the internal structure of HMCs as observed in Figure~\ref{fig6}d.

Assuming that the 3~mm flux density is due to warm dust, we derive a clump mass M$_{clump}$~=~$g S_{\nu} d^2$~/~$\kappa_{\nu} B_{\nu}(T_d)$ \citep[e.g.,][]{B06} for the G12--HMC of $\sim$~2.0$\times$10$^{3}$~M$_{\sun}$, where a gas--to--dust ratio of g $=$ 100 and a dust mass opacity coefficient of $\kappa_{112}~=~0.2~cm^2~g^{-1}$ \citep{OH94} were adopted. A flux density of 0.15~Jy (Table~\ref{vla_cont}), a distance of 13.5~kpc, and a temperature T$_d$~=~86~K (Table~\ref{hotcore}). In addition, considering a 3~mm deconvolved angular size of 1.7$\pm$0.5$''$, and assuming that all hydrogen is in molecular form and the clump is spherical, we determine a n(H$_{\rm 2}$)$_{\rm 3mm}$~$\sim$~6.3$\times$10$^6$~cm$^{-3}$, and thus a central hydrogen column density of N(H$_{\rm 2}$)~$\sim$~4.2$\times$10$^{24}$~cm$^{-2}$ \citep[following e.g.,][]{S-M17}. Both values of n(H$_{\rm 2}$) and N(H$_{\rm 2}$) are in agreement with those reported in Table~\ref{previous}.

The gas mass can be estimated from the $^{13}$CS(2$-$1) integrated line flux density assuming LTE, optically thin emission, and a fractional abundance for the $^{13}$CS molecule. Then, assuming the $^{12}$CS abundance is CS/H$_2 =10^{-8}$, as estimated for the Orion Hot Core \citep{VB98}, $^{12}$CS/$^{13}$CS ratio of 40 \citep{T73}, and excitation temperature $\simeq$~86~K from NH$_{\rm 3}$ data, we obtain a gas mass of ~6.3$\times$10$^3$~M$_{\sun}$. Then, the mass ratio for G12--HMC is M($^{\rm 13}$CS)/M(H$_{\rm 2}$)~$\sim$~3. This discrepancy in the estimated mass can be explained in terms of the uncertainties in the adopted abundances, on the hot dust emission law and the gas to dust ratio. 

Under the consideration of uniform density, a $^{\rm 13}$CS(2$-$1) virial mass of $\sim$2200~M$_{\sun}$ is derived. This virial mass is in a better agreement with the NH$_{\rm 3}$ virial mass ($\sim$1500~M$_{\sun}$ and $\sim$1800~M$_{\sun}$) in comparison with that determined by \citet{H98} via JCMT observations of CH$_{\rm 3}$OH (1300~M$_{\sun}$). 

The volumetric density of H$_{\rm 2}$ was also determined using equation 2 of \citet{HT83}. Considering the Einstein coefficient A(s$^{-1}$) for the NH$_{\rm 3}$(2,2) transition, the collision rate C(s$^{-1}$) from \citet{D88}, we obtain n(H$_{\rm 2}$)~$\sim$~1.5$\times$10$^6$~cm$^{-3}$. The total NH$_{\rm 3}$ column density was determined by summation of N(J,K) over all metastable rotational levels up to level J~=~5, using Boltzmann's Equation and assuming LTE. This density is somewhat less than that calculated using the 3~mm emission ($\sim$6.3$\times$10$^6$cm$^{-3}$).

In addition, the H$_{\rm 2}$ column density was calculated as N(H$_{\rm 2}$)~=~${\rm \Theta_S}$~n(H$_{\rm 2}$) where ${\rm \Theta_S}$ is the deconvolved linear size in cm ($3\rlap{.}{''}4 \pm$0.1 $\sim$~0.22$\pm$0.01~pc at 13.5~kpc; using ${\rm \Theta_S}$~=~$\sqrt{\theta_x \theta_y}$ from Table~\ref{nh3obs}). The relative abundance (X$_{\rm NH_3}$) between NH$_{\rm 3}$ and H$_{\rm 2}$ was calculated as N(NH$_{\rm 3}$)/N(H$_{\rm 2}$). A molecular gas mass of the clump using ammonia, M(NH$_{{\rm 3}}$), was determined using equation 12 of \citet{GL99} with a radius R~=~0.11$\pm$0.01~pc, giving a value of M(NH$_{\rm 3}$)~=~830$\pm$40~M$_\odot$. 

Since the virial mass estimations (2200~M$_{\sun}$ from $^{13}$CS(2$-$1), and $\sim$1500~M$_{\sun}$ to $\sim$1800~M$_{\sun}$ from NH$_{\rm 3}$) are about twice the ammonia gas mass ($\sim$~830~$\pm$40~M$_{\sun}$) we conclude that the G12--HMC may not be gravitationally bound. The physical parameters of this HMC are summarized in Table~\ref{hotcore}.

\subsection{HMC Internal Heating Source}
\label{disnh3}

From the flux density at 3~mm and the upper limit at 7~mm from the molecular clump emission (3$\sigma$~=~1.8~mJy), a spectral index of $\alpha\gax$4.5 is obtained. This value suggests that the 3~mm emission arises from optically thin dust. If we consider that the dust in the molecular clump is only heated by external radiation, a lower limit to the dust temperature can be obtained using the formulation of \citet[equation 9]{SK76}, under the assumption that no absorption exists between the ionizing star of the \uchii region and the clump. Using the 3~mm integrated flux density (see Table~\ref{vla_cont}) and assuming an external O6.5~ZAMS ionizing star \citep[L$_{\rm {star}}$~$\sim$~1.5$\times$10$^5$L$_{\rm {\sun}}$]{KK01}, a dust absorptivity $\kappa_\nu \propto \nu^{+2}$, and a projected separation between the clump and the ionizing star (which is presumed to be at the focus of the cometary \uchii region) of 0.26~pc (4$\asec$ at 13.5~kpc), we obtain T$_{\rm d}$~=~130~K. This temperature is slightly greater than the values obtained with other molecular tracers (see Table~\ref{previous}, \S~\ref{nh3} and \S~\ref{disnh3}). However, not all luminosity of the O6.5~ZAMS ionizing star reaches the HMC. Considering a dilution factor of $\Omega$/4$\pi$~$\sim$0.06 (where $\Omega$ is the solid angle subtended by the ammonia core as seen from the ionizing star of the \uchii),the fraction of incident luminosity from the \uchii on the HMC would be L$_{\rm inc}$~$\sim$~9000~L$_{\rm {\sun}}$. Using this quantity as the energy budget of the HMC, a dust temperature of about 70~K is estimated, which is similar to the kinetic temperature obtained in section \S~\ref{nh3} (86$\pm$12~K). This implies that an external heating source is not needed to explain the temperature of the core. However, the detection of water and methanol masers suggests the presence of a YSO in the HMC.

\subsection{HMC Kinematics}
\label{disKin}

There are different types of gas motion that can ocurr in HMCs, i.e., infall, rotation, and expansion/outflows. Outflows and rotation have been invoked to explain velocity gradients toward several HMCs \citep{C98}. In the case of rotation, flattened compact `disk--like' structures with diameters greater than 0.1~pc \citep[][and references therein]{GL99} are observed, and the velocity gradient may indicate the kind of rotation: differential or rigid body.

\begin{figure}
\includegraphics[width=\columnwidth]{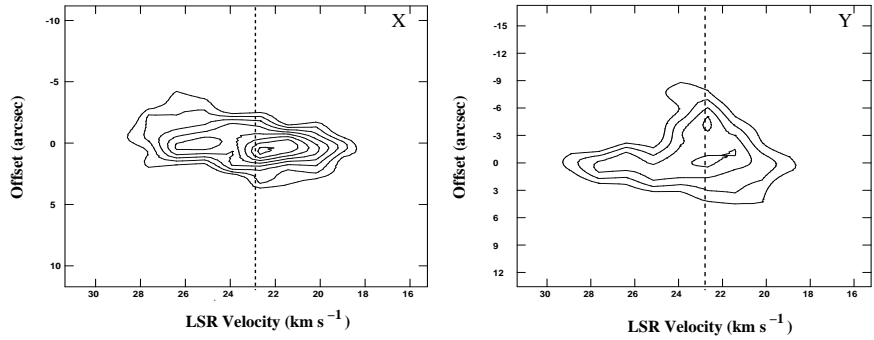}
  \caption{NH$_{\rm 3}$(2,2) position--velocity (PV) diagram of the HMC presented as grey-scale in Figure~\ref{fig6}a and \ref{fig6}b. The plots are slices in the directions X and Y shown in the lower-right corner in Figure~\ref{fig6}b. The origin of the position axes is located at $\alpha_{2000}$~=~18$^{\rm h}$~12$^{\rm m}$~40$^{\rm s}$, $\delta_{2000}$~=~--18$^{\circ}$~24$'$~15$''$.  Contours are --4, 4, 5, 6, 7, 8, 9, 10, 11 $\times$ 1.6~mJy~beam$^{-1}$. Dashed lines mark the systemic velocity of the core at $\sim$23\kms.} 
 \label{fig7}
\end{figure}

The NH$_{\rm 3}$(2,2) position--velocity (PV) diagrams of G12--HMC are presented in Figure~\ref{fig7}. We find no evidence of infall (e.g., redshifted absorption features). The central parts of the G12--HMC present larger velocity widths compared to the edges ($\sim$6 versus 2\kms), and the overall structure of the HMC is elongated. The rotation scenario is possible. 

The velocity widths across the HMC in the PV diagrams are also consistent with expansion motions. Two components on each side of the HMC central position(systemic velocity of $\sim$23\kms), have velocities of $\sim$26\kms ~and $\sim$20\kms, suggesting that G12--HMC could be a structure with expansion motions at velocities of a few~km~s$^{-1}$ in the SE--NW direction.

We have obtained images of integrated velocities from $\sim$30 to 23\kms ~and from 23 to 18\kms ~using only the main line, and we do not see a clear spatial separation between this red--shifted and blue--shifted components. The opposite was found in the high--mass star forming regions G25.65+1.05 and G240.31+0.07, where high velocity molecular gas is associated with bipolar outflows \citep{SC96}. Higher angular resolution observations are needed to confirm the rotation hypothesis and whether an outflow is present in the core.

\section{Summary and Conclusions}
\label{summary}

We present observations of radio continuum (0.3, 0.7, 2 and 3.6~cm) and spectral lines (H41$\alpha$, $^{13}$CS(2$-$1) and (1$-$0), NH$_{\rm 3}$(2,2) and (4,4)) toward the cometary \hii region G12.21--0.10. Continuum emission  is detected at all wavelengths toward the UC~H~II region. We also find 3~mm continuum emission toward a molecular clump located about 4$''$ from the UC~H~II region. $^{\rm 13}$CS and NH$_{\rm 3}$ emission is only detected toward the molecular clump and we do not find any chemical differentiation between these species.

We find evidence that the molecular clump is consistent with a hot (86~K), dense (1.5$\times$10$^6$ cm$^{-3}$), big (0.22~pc), and turbulent ($\rm \Delta V_{\rm obs} \sim$~8~\kms) molecular core. We also find that this hot molecular core shows a marginal velocity gradient in the SE--NW direction. Given that the HMC coincides with a water and methanol maser group, we speculate that a YSO, not detected in the centimeter radio continuum, could be forming inside the HMC.

Finally, we find that $^{\rm 13}$CS and NH$_{\rm 3}$ are good candidates to trace the internal structure of HMCs, however, high angular resolution and higher sensitive observations are needed to fully characterize this HMC.

\acknowledgments

The authors thank the anonymous referee for useful comments which improved this work. E. F. very gratefully acknowledges financial support from CONACyT (grant 124449, SNI III 1326 M\'exico), PROMEP$/$103.5$/$08$/$4722, and facilities from NRAO, NMT, IRyA--UNAM, INAOE. He also acknowledges Benem\'erita Universidad Aut\'onoma de Puebla (BUAP); Arturo Fern\'andez and Humberto Salazar, for several supports during a research stay on December, 2017. We would like to thank Laurant Loinard, Mayra Osorio, Guillem Anglada, Jana Benda, Simon Kemp, Susana Lizano, and Luis Felipe Rodr\'{\i}guez for useful discussions. We also thank Gilberto Zavala, Miguel Espejel, Alfonso Ginori and Andr\'es Rodr\'\i guez (IAM--UdeG) for computer support. P. H. acknowledges partial support from NSF grant AST-0908901 for this project.

\begin{table}[!t]\centering
  \setlength{\tabnotewidth}{1.0\columnwidth}
  \tablecols{6}
  \caption{Previous Molecular Observations Toward G12} \label{previous}
\begin{tabular}{cccccc}\toprule
Molecule & T$_{\rm k}$  & n(H$_{\rm 2}$) & Size     &  N(H$_{\rm 2}$) & Reference\tabnotemark{a} \\
         & (K)          & (cm$^{-3}$)    & (arcsec) & (cm$^{-2}$)     & \\  \midrule
NH$_{\rm 3}$ & \nodata  & 4.8$\times$10$^5$ & 4.0     & 5.2$\times$10$^{23}$      & (1)  \\
CS       & \nodata      & 8.1$\times$10$^5$ & \nodata & 2.0$\times$10$^{23}$      & (2),(5)  \\
CH$_{\rm 3}$CN & 80--90 & 10$^6$--10$^9$    & \nodata & 2.0$\times$10$^{25}$      & (3),(1)  \\
CH$_{\rm 3}$OH & 56${{+800}\atop{-24}}$ & 10$^7$ & 1.3 & 1.2 $\times$ 10$^{25}$ & (4) \\  \bottomrule
\tabnotetext{a}{(1) Cesaroni et al. 1992; (2) Plume et al. 1997; (3) Olmi et al. 1993; (4) Hatchell et al. 1998, (5) Shirley et al. 2003.}
\end{tabular}
\end{table}

\begin{table}\centering
  \setlength{\tabnotewidth}{1.0\columnwidth}
  \tablecols{5}
\caption{Detection Summary} \label{pos}
\scriptsize
\begin{tabular}{ccccc}\toprule
Observation & Wavelength & Right Ascencion & Declination & Source \\
            & {(cm)}     & (J2000) & (J2000) &  \\  \midrule
 VLA continuum & 0.7, 2, \& 3.6  & 18$^{h}$~12$^{m}$~39.8$^{s}$ & --18$^{\circ}$~24$'$~21$''$  & \uchii region  \\ 
OVRO H41$\alpha$         & 0.33  & 18$^{h}$~12$^{m}$~39.6$^{s}$ & --18$^{\circ}$~24$'$~21$''$  & \uchii region \\
OVRO continuum           & 0.30  & 18$^{h}$~12$^{m}$~39.8$^{s}$ & --18$^{\circ}$~24$'$~21$''$  & \uchii region (peak--1)\\ 
OVRO continuum           & 0.30  & 18$^{h}$~12$^{m}$~39.8$^{s}$ & --18$^{\circ}$~24$'$~17$''$  & molecular clump (peak--2)\\ 
OVRO $^{13}$CS(2$-$1)    & 0.32  & 18$^{h}$~12$^{m}$~39.7$^{s}$ & --18$^{\circ}$~24$'$~18$''$  & molecular clump  \\
VLA NH$_3$(2--2)         & 1.26  & 18$^{h}$~12$^{m}$~39.8$^{s}$ & --18$^{\circ}$~24$'$~18$''$  & molecular clump  \\
VLA NH$_3$(4--4)         & 1.24  & 18$^{h}$~12$^{m}$~39.8$^{s}$ & --18$^{\circ}$~24$'$~18$''$  & molecular clump  \\
VLA $^{13}$CS(1$-$0)     & 0.65  & 18$^{h}$~12$^{m}$~39.8$^{s}$ & --18$^{\circ}$~24$'$~18$''$  & molecular clump  \\
VLA continuum\tabnotemark{a} & 1.3   & 18$^{h}$~12$^{m}$~39.7$^{s}$ & --18$^{\circ}$~24$'$~21$''$  & \uchii region  \\  \bottomrule
\tabnotetext{a}{This is the continuum obtained from NH$_3$ observations.}
\end{tabular}
\end{table}

\begin{table}\centering
  \setlength{\tabnotewidth}{1.0\columnwidth}
  \tablecols{7}
\caption{Continuum Parameters} \label{vla_cont}
\scriptsize
\begin{tabular}{ccccccc}\toprule
 Region & Telescope  &  $\lambda$ &  $S_\nu$ &  Beam Size; P.A  &   RMS Noise &  Deconvolved Size \\
  &   &  (cm) &  (mJy)\tabnotemark{a} &  (arcsec; deg) &  (mJy beam$^{-1}$)  &  (arcsec) \\ \midrule
\uchii& VLA CnB & 3.6 & 220 & 2.31 $\times$ 0.91; $-70$ & 0.31 & 5 $\times$ 4  \\
      & VLA CnB & 2.0 & 200 & 1.04 $\times$ 0.55; $+80$ & 0.21 & 8 $\times$ 7  \\
      & VLA CnB & 0.7 & 160 & 1.65 $\times$ 0.83; $+77$ & 0.58 & 4 $\times$ 5  \\  
Molecular & &&&&&  \\  
 clump & OVRO & 0.3 & 150 &2.82 $\times$ 2.32; $-43$ & 1.70  & 3 $\times$ 1  \\
\\
Extended & &&&&&  \\ 
emission & VLA D & 3.6 & 1320 &12.50 $\times$ 7.30; $-13$ & 0.15 & 210 $\times$ 156  \\   \bottomrule
\tabnotetext{a}{Uncertainties in the integrated flux densities are estimated to be about 5\% at 3.6, 10\% at 2~cm, and 20\% at 7~mm. The deconvolved sizes and uncertainties in flux densities were obtained using the task IMFIT in AIPS.}
\end{tabular}
\end{table}

\begin{table}\centering
  \setlength{\tabnotewidth}{1.0\columnwidth}
  \tablecols{7}
\caption{$^{13}$CS(2$-$1) and H41$\alpha$ results} \label{h41cspar}
\scriptsize
\begin{tabular}{ccccccc}\toprule
 Line\tabnotemark{a} &  Beam Size; P.A. &  RMS noise &  S$_\nu$\tabnotemark{b}  &  V$_{\rm LSR}$ &  $\Delta$V$_{\rm obs}$ &  Deconvolved Size\tabnotemark{c}  \\
&  (arcsec; deg) &  (mJy beam$^{-1}$) &  (mJy) & km s$^{-1}$ &  km s$^{-1}$ &  (arcsec) \\ \midrule
$^{13}$CS(2-1) & 3.18 $\times$ 2.51; $-43$ & 35 & 300  &  23.7$\pm$0.1  & 9.4$\pm$0.5 &  4.5 $\times$ 2.0 \\
H41$\alpha$ & 3.18 $\times$ 2.51; $-43$  & 35 & 335 & 29.4$\pm$1.2  & 23.9$\pm$3.1 & \nodata  \\   \bottomrule
\tabnotetext{a}{The spectral resolution of the CS line is 1.6 km~s$^{-1}$ and that of the H41$\alpha$ line is 3.3 km s$^{-1}$.}
\tabnotetext{b}{From peak channel at 23\kms ~for $^{13}$CS(2$-$1), and at 30\kms ~for H41$\alpha$.}
\tabnotetext{c}{The size of the source was obtained from the moment 0 image using IMFIT in AIPS.}
\end{tabular}
\end{table}

\begin{table}\centering
  \setlength{\tabnotewidth}{1.0\columnwidth}
  \tablecols{7}
\caption{NH$_3$ Line Parameters} \label{nh3obs}
\scriptsize
\begin{tabular}{ccccccc}\toprule
 Transition &  Beam Size; P.A. &  RMS Noise\tabnotemark{a} &  Peak S$_\nu$ &  V$_ {\rm {LSR}}$ &  $\Delta {\rm V_{obs}}$ &  Deconvolved Size \\
 (J,K; line) &  (arcsec; deg) &  (mJy beam$^{-1}$) &  (mJy) &  (km s$^{-1}$) &  (km s$^{-1}$) &  (arcsec)\\ \midrule
(2,2; main) & 3.54$\times$1.36; $+60$ & 4.4 & 260$\pm$5 & 23.1$\pm$0.3\tabnotemark{a} & 8.1$\pm$0.7 & 3.8 $\times$ 3.0  \\
(2,2; inner satellites) & 3.54$\times$1.36; $+60$ & 5.7 & 59$\pm$2 & 6.5, 39.7\tabnotemark{b} & 9.5\tabnotemark{c} & \nodata \\
(2,2; outer satellites) & 3.54$\times$1.36; $+60$ & 6.1 & 59$\pm$2 & --2.7, 48.9\tabnotemark{b} & 8.5\tabnotemark{c} & \nodata \\
(4,4; main) & 4.29$\times$1.18; $+56$ & 3.2 & 106$\pm$5 & 23.3$\pm$0.3\tabnotemark{a} & 8.9$\pm$0.7 & 3.0 $\times$ 1.0 \\
(4,4; inner satellites) & 4.29$\times$1.18; $+56$ & 3.3 & 16.0$\pm$0.8 & --0.7, 27.3\tabnotemark{b} & 13.2\tabnotemark{c} & \nodata \\
(4,4; outer satellites) & 4.29$\times$1.18; $+56$ & 3.3 & 16.0$\pm$0.8 & --6.7, 53.3\tabnotemark{b} & \nodata & \nodata \\   \bottomrule


\tabnotetext{a}{Uncertainties are from Gaussian fits. The flux densities are from a box covering the HMC.}
\tabnotetext{b}{These two velocity values correspond to satellite lines without fitting.}
\tabnotetext{c}{Average values between the corresponding pair from the gaussian fitting.}
\end{tabular}
\end{table}

\begin{table}\centering
  \setlength{\tabnotewidth}{1.0\columnwidth}
  \tablecols{6}
\caption{Physical parameters of the Hot Molecular Core} \label{hotcore}
\begin{tabular}{cccccc}\toprule
 T$_{42}$ \tabnotemark{a} &  T$_{\rm k}$ &  N(H$_{\rm 2}$) &  M(NH$_{\rm 3}$) &  N(NH$_{\rm 3}$) &  X(NH$_{\rm 3}$) \\
 (K) &  (K) & (10$^{\rm 24}$ cm$^{\rm -2}$) &  (M$_\odot$) &   (10$^{\rm 17}$ cm$^{\rm -2}$) &  (10$^{\rm -7}$) \\ \midrule
$77\pm10$ & $86\pm12$ & 1.0$\pm$0.4 & 830$\pm$40 & $1.1\pm0.4$ & 1.1$\pm$0.7 \\  \bottomrule
\tabnotetext{a}{Rotational temperature estimated from NH$_3$(2,2) \& (4,4) observations.}
\end{tabular}
\end{table}

\end{document}